\documentclass{ws-procs9x6}

\def \pt      {p_T}

\def \snn     {\sqrt{s_{NN}}}

\def \pbar    {\overline{p}}

\def \ncollav {\langle N_{coll} \rangle}

\def \propp     {p/\pi^{+}}
\def \pbarpm  {\pbar/\pi^{-}}

\def \pm  {\pi^{-}}

\begin{document}

\title{Parton energy loss, saturation, and recombination at BRAHMS}

\author{Eun-Joo Kim for the BRAHMS Collaboration}
\address{University of Kansas, \\ 
Lawrence, Kansas 66045, USA \\
E-mail: ejkim@ku.edu}  

\maketitle

\abstracts{
Particle production as observed with the BRAHMS experiment 
at RHIC is presented. Preliminary baryon/meson ratios 
and nuclear modification factors at different rapidities
will be discussed.
}

\section{Introduction}

Hadrons with high transverse momentum provide a good probe of 
the high energy density matter created at RHIC, 
since the production of high $\pt$ particles is dominated 
by the initial hard parton-parton scatterings with large momentum transfer $Q^{2}$. 
After hard-scattering, partons traverse a medium with a high density 
of color charges where they interact strongly, emit gluon radiation,
and lose energy before fragmenting into hadrons.
The production of hadrons depends on the initial parton distributions
in the colliding nuclei, the elementary parton-parton cross section 
and the hadronization process of partons into hadrons.
It is also important to distinguish nuclear effects from initial state effects, 
such as described by shadowing and/or color glass condensate models, and final state effects.
To disentangle all these behaviors requires a very comprehensive data set. 
The BRAHMS experiment\cite{brahms_nim,brahms_white} has studied
p+p, d+Au, and Au+Au collisions over a broad range of rapidity and transverse momentum. 
We will discuss these data in the context of the above processes.

\section{Result}

High $\pt$ suppressions have been observed 
in central Au+Au collisions at RHIC\cite{phenix,star,phobos} 
and are attributed to final-state interactions 
based on the absence of such suppressions 
in d+Au collisions\cite{brahms_highpt,phenix_da,star_da,phobos_da}.
The suppression is quantified by use of nuclear modification factors,
which are defined as $R_{AA}$ or $R_{CP}$ :

\begin{equation}
R_{AA} \equiv \frac{1}{\langle N_{coll} \rangle}
        \frac{d^2N^{AA}/dp_Tdy}{d^2N^{pp}_{inel}/dp_Tdy}, 
R_{CP} \equiv \frac{\frac{1}{\langle N^{C}_{coll} \rangle}{d^2N^{C}/dp_Tdy}}
                   {\frac{1}{\langle N^{P}_{coll} \rangle}{d^2N^{P}/dp_Tdy}}
\label{eq:NMF}
\end{equation}

$R_{AA}$ gives the deviation in yields from AA collisions 
relative to the scaled yields from nucleon-nucleon collisions.
$R_{CP}$ can provide similar information 
based on the relative yield in central(C) and peripheral(P) collisions
scaled by the mean number of binary collisions, 
but does not depend on the reference nucleon-nucleon system.

Figure~1 shows the rapidity(a) and particle dependence(b) 
of $R_{CP}$ in Au+Au collisions at $\snn = $ 200~GeV.
The observed suppression is similar 
at forward rapidities~($\eta \sim$ 2.2, 3.2) as compared to midrapidity. 
This result may indicate quenching extends in the longitudinal direction.
$R_{CP}$ for protons reaches unity  around $\pt \sim$ 1.5~GeV/$c$, 
but $R_{CP}$ for pions is suppressed at higher $\pt$.
The difference between baryon and meson behaviors is discussed later. 

\begin{figure}[ht]
\centerline{\epsfxsize=4.1in\epsfbox{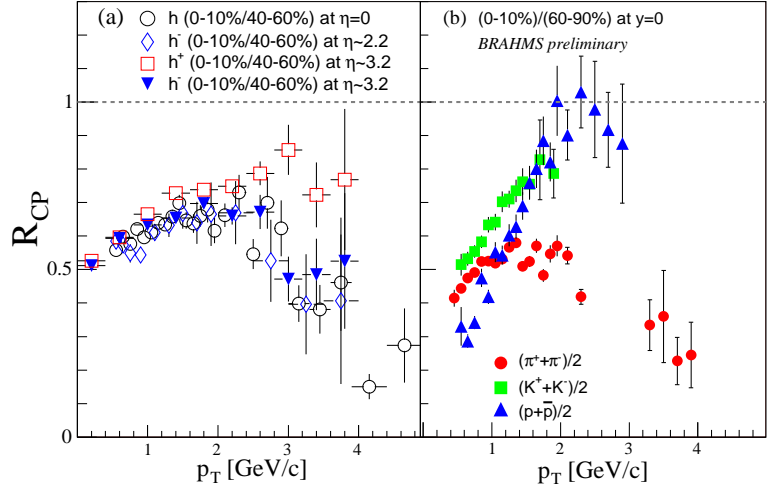}}   
\caption{(a) Nuclear modification factor for the most central and 
peripheral collisions at pseudorapidities $\eta = 0,~2.2, ~3.2$. 
The values for $\eta = 0,~2.2$ are from BRAHMS publication$^{6}$,
and the one for $\eta = $3.2 is preliminary result. 
(b) Central (0-10\%) to peripheral (60-90\%) ratios, $R_{CP}$, 
as a function of $\pt$ for identified hadrons at midrapidity. 
(a) and (b) are from Au+Au collisions at $\snn = $200~GeV. 
Error bars are statistical only.}
\end{figure}

\begin{figure}[ht]
\centerline{\epsfxsize=5.12in\epsfbox{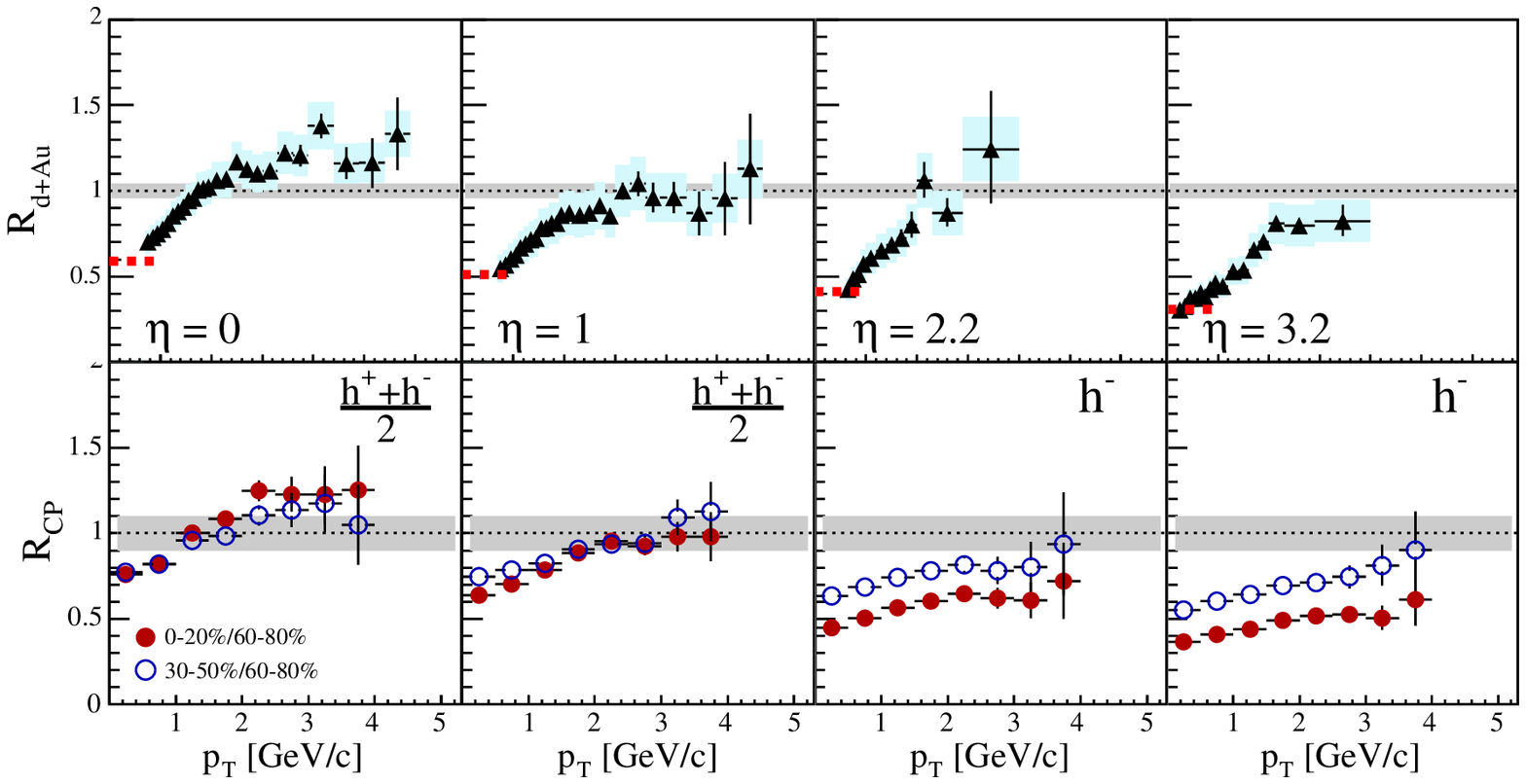}}   
\caption{Top row : Nuclear modification factor for charged hadrons at
pseudorapidities $\eta = 0,~1.0,~2.2,~3.2$. 
Systematic errors are shown with shaded boxes with widths set by the bin sizes.
Bottom row : Central(field circles) and semi-central(empty circles) $R_{CP}$
ratios in d+Au collisions at $\snn = $ 200~GeV. 
Shaded bands indicate the uncertainty in the calculation of $\ncollav$
in the peripheral collisions~(12\%).}
\end{figure}

The rapidity dependence of $R_{dA}$ and $R_{CP}$ 
for d+Au collisions~\cite{brahms_rda} is shown in Fig.~2.
At midrapidity, $R_{dA}$($\pt > $ 2~GeV/$c$) shows a Cronin type
enhancement compared to the binary scaling limit.
At higher rapidity, this enhancement is followed by
a suppression which becomes stronger at forward rapidity.
Along the bottom row, the $R_{CP}$ for two different centrality ranges
is shown as function of pseudorapidity.
The more central $R_{CP}$ exhibits greater suppression as the rapidity increases.
This is consistent with the picture of parton saturation 
in the Au-wave function\cite{cgc_da}.
However, the suppression of $R_{CP}$ at forward rapidity can also be reproduced 
in the framework of parton recombination in the final state\cite{reco_da},
without involving multiple scattering and gluon saturation in the initial state.

\begin{figure}[ht]
\centerline{\epsfxsize=4.1in\epsfbox{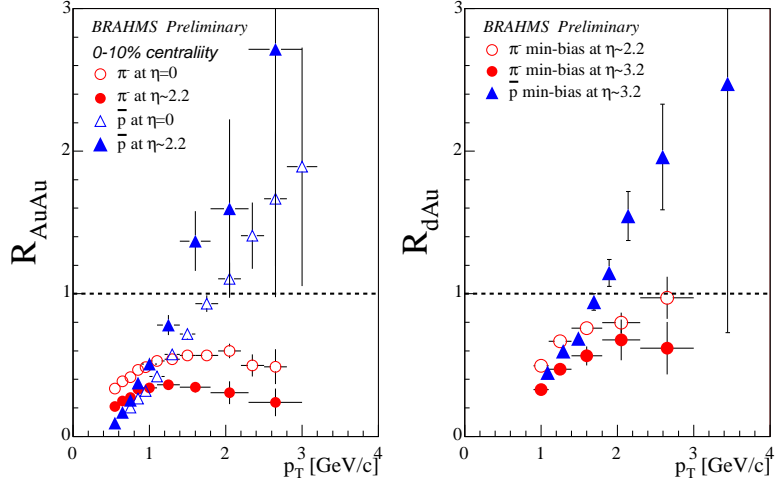}}   
\caption{(left panel) $R_{AuAu}$ for $\pm$ and $\pbar$ 
at midrapidity and forward rapidity
for 0-10\% central Au+Au collisions at $\snn = $ 200~GeV.
(right panel) $R_{dAu}$ of $\pm$ and $\pbar$ at forward rapidity, 
$\eta =$ 2.2 and 3.2 for d+Au collisions at $\snn = $ 200~GeV. 
No weak decay feed-down correction applied.}
\end{figure}

Figure~3 shows the dependence of the high $\pt$ behavior 
on the type of particle in d+Au and Au+Au collisions.
Results in Au+Au collisions show $\pm$ are suppressed 
at midrapidity and forward rapidity.
At forward rapidity, the suppression is stronger for $\pm$,
while the $\pbar$ yields are enhanced at both rapidities.
In d+Au collisions, the $\pm$ yields are more suppressed at $\eta \sim$ 3.2, 
while, again, the $\pbar$ yields are enhanced at forward $\eta$.
This different behavior between $\pm$ and $\pbar$ is not 
consistent with standard fragmentation functions, 
and indicates pions experience high $\pt$ suppression
while protons do not. This is not yet fully understood.
Proton excess might arise from hydrodynamic expansion, or
parton recombination\cite{reco_ratio} 
and/or quark coalescence\cite{coal_ratio}
processes that enhance the yield of baryons containing three quarks 
by pulling them from the medium rather than 
relying on a simple fragmentation origin.  

The measured $\propp$ and $\pbarpm$ ratios as a function of $\pt$ 
for central Au+Au collisions at different rapidities are shown in Fig.~4.
There is a clear increase of the $p/\pi$ ratios 
at intermediate $\pt$ ($2< \pt < 5$~GeV/$c$)
relative to the level seen in nucleon-nucleon collisions\cite{ratioinpp,ratioinee}. 
There is no significant difference for the ratios 
at rapidity $y=0$ and $y \sim 1$, 
and  $\pbarpm$ ratio shows a similar tendency up 
to $\pt \sim 1.5$ GeV/$c$ at $\eta \sim 2.2$.

\begin{figure}[ht]
\centerline{\epsfxsize=3.91in\epsfbox{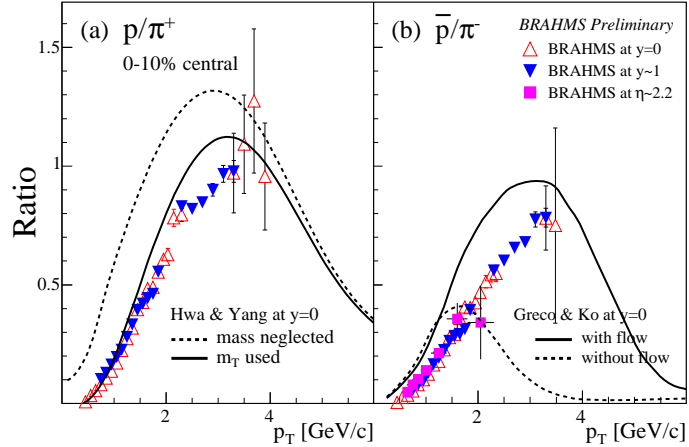}}   
\caption{$\propp$ (a) and $\pbarpm$ (b) ratios 
at rapidity $y = 0,~ 1.0 $ and $\eta = 2.2$.  
for 0-10\% central Au+Au collisions at $\snn = $ 200~GeV.
Feed-down corrections applied.
Comparisons with model calculations$^{13,14}$ are shown.}
\end{figure}

\section{Summary}

BRAHMS has measured rapidity dependent nuclear modification factors
and particle ratios in different colliding systems. 
The evolution of nuclear modification factors in d+Au collisions 
may indicate parton saturation in the initial state.
The high $\pt$ suppression in Au+Au collisions at midrapidity
also exists at forward rapidity, and depends on particle type.
The recombination/coalescence models seem to give a reasonable
explanation of the observed baryon-meson production mechanism
at intermediate $\pt$.

\section{Acknowledgments}

This work was supported by the division of Nuclear Physics of the
Office of Science of the U.S. DOE, the Danish Natural Science
Research Council, the Research Council of Norway, the Polish State
Committee for Scientific Research and the Romanian Ministry
of Education and Research.

\end{document}